# What probabilities tell about quantum systems, with application to entropy and entanglement

John M. Myers

Division of Engineering and Applied Sciences, Harvard University,

60 Oxford Street, Cambridge, MA 02138, USA

F. Hadi Madjid

82 Powers Road, Concord, MA 01742, USA

**Abstract**

As described quantum mechanically, an experimental trial parses into "a preparation" expressed by a density operator and "a measurement" expressed by a set of detection operators, one for each measurable event. A density operator and a detection operator combine via the "trace rule" to generate the probability of a measurable event. As used to describe experiments, both density operators and detection operators depend on parameters expressing experimental choices, so the probabilities they generate also depend on these parameters. The trace rule answers the question: "what parametrized probabilities are generated by a given parametrized density operator and given parametrized detection operator?"

In comparing an experiment against its description by a density operator and detection operators, one compares tallies of experimental outcomes against the probabilities generated by the operators but not directly against the operators. Recognizing that the accessibility of operators to experimental tests is only indirect, via probabilities, motivates us to ask what probabilities tell about operators, or, put more precisely: "what combinations of a parametrized density operator and parametrized detection operators generate any given set of parametrized probabilities?"

We review and augment recent proofs that any given parametrized probabilities can be generated in very diverse ways, so that a parametrized probability measure, detached from any of the (infinitely many) parametrized operators that generate it, becomes an interesting object in its own right. By detaching a parametrized probability measure from the operators that may have led us to it, we (1) strengthen Holevo's bound on a quantum communication channel and (2) clarify a role for multiple levels of modeling in an example based on quantum key distribution. We then inquire into some parametrized probability measures generated by entangled states and into the topology of the associated parameter spaces; in particular we display some previously overlooked topological features of level sets of these probability measures.

## 1. INTRODUCTION

Familiar in quantum physics is the use of parameters to describe an experimenter's control over the devices used in an experiment, for example in connection with Bell inequalities. Recently we became interested in parameters in a different but related context, when we proved a formal separation in quantum mechanics between linear operators and the probabilities that these operators generate.

As described quantum mechanically, an actual or contemplated experimental trial parses into: (1) "a preparation" expressed by a density operator (for a pure or mixed quantum state) and (2) "a measurement" expressed by the detection operators associated with a positive operator-valued measure (POVM) [1–3]. By a convention of quantum physics, one combines a density operator $\rho$ and a detection operator $M(\omega)$ in the "trace" formula by which these two generate a probability $\mu(\omega)$ for a measurable event $\omega$:

$$\mu(\omega) = \text{Tr}[\rho M(\omega)]. \tag{1.1}$$

(A special case is that of pure states (wave functions) $|\psi\rangle$ for which the density operator has the form $\rho = |\psi\rangle\langle\psi|$; on the POVM side, a special case is that in which the detection operators are projections.)

In comparing an experiment against its description by a density operator and detection operators, one compares tallies of experimental outcomes [4] not against the operators directly but against the probabilities calculated from the density operator and the detection operators per Eq. (1.1). Recognizing that the operators are tested experimentally only indirectly by way of probabilities, we wondered: what can given probabilities tell us about density operators and detection operators? (To begin with, we ask this question without the "tensor-product" restrictions imposed on the detection operators in some prior investigations [5].)

A little thought shows that any given probability $\mu(\omega)$ can be generated by distinct choices of $\rho$ and $M(\omega)$, but we wanted to know how the dependence of $\mu$, $\rho$ and $M$ on experimental control affects this freedom of choice. Characterizing control over an experiment by some list $k$ of parameter values, we introduced a set $\mathcal{K}$ of lists of parameter values as the domain of three functions:

1. a function that assigns to each parameter list $k \in \mathcal{K}$ a density operator $\rho^{(k)}$;

2. a function that assigns to each $k$ a POVM $M^{(k)}$; and

3. a function that assigns to each $k$ a probability measure $\mu^{(k)}$.

In partial answer to our question, we proved that no assignment of probability measures to parameter lists can ever uniquely determine the density operators and POVMs [6]. This proof reveals a logical gap between parametrized probabilities and the parametrized operators that generate them, thereby making parametrized probability measures objects of interest in their own right, as will be demonstrated in the following sections. In Secs. 3 A and 3 B, we give two examples of uses of parametrized probability measures independent of the particular quantum states and POVMs that led us to them. The first example pertains to the information capacity of a quantum channel, while the second pertains to quantum cryptography.

In Sec. 4 we study parametrized probability measures associated with some entangled states. While pure entangled states violate Bell inequalities [7–9], these inequalities pertain directly not to quantum states as density operators but to the parametrized probability measures ensuing from these states conjoined to parametrized POVMs. We inquire into the topology of the associated parameter spaces and we display topological features of level sets of the probability measures as functions on the parameter space. Some open questions are sketched. In the next section we begin with definitions, some probably novel, others previously known but perhaps unfamiliar.

## 2. PARAMETER SPACES AND QUANTUM MODELS

In setting up an experiment, say on an optics bench, one reaches not for a density operator or a POVM, but for lenses and lasers. Yet neither 'lens' nor 'laser' is a term in the language of quantum mechanics. Indeed if the devices of an experiment—lasers, detectors, filters, and so on—were fixed, a quantum theorist would abstract the devices used in the experiment into a single point to which he or she attaches a density operator and a POVM. Generally, however, the experimenter adjusts the devices: by turning one knob to rotate a polarizing filter, another knob to change the position of a lens, another to control the temperature of a crystal, *etc*. Correspondingly, the experiment tallies detections not just for a run of trials corresponding to a single value of a parameter list, but for runs corresponding to a variety of values. We express the possible positions of the "knobs" for some particular experiment, actual or contemplated, by a set $\mathcal{K}$; for some fixed positive integer $\ell$, any element $k \in \mathcal{K}$ is a list $\{k_1, k_2, \ldots, k_\ell\}$ in which $k_j$ can be either a continuous (real-valued) parameter or a discrete parameter. In most cases there is a topology on $\mathcal{K}$, in which case we speak of a parameter *space*. We will see an example in which $\mathcal{K}$ is a torus. On behalf of simplicity, here we restrict attention to experimental parameters that, while they can change in value from trial to trial, are held constant during any one trial. (In contrast, Berry phases arise from modeling what happens to a quantum state when parameters are changed during a trial [10].)

An experimenter can have occasion to add a device or to replace a device with one that works over a wider range of parameter values. Such an extension of the experiment calls for extending its description, including extending the set $\mathcal{K}$ into a bigger set $\mathcal{K}'$ via one or both of two types of injective mapping $\Xi : \mathcal{K} \hookrightarrow \mathcal{K}'$:

1. $\mathcal{K}'$ increases the range of one or more of the parameters of the list in $\mathcal{K}$, so that $\mathcal{K} \subset \mathcal{K}'$; and/or

2. the lists in $\mathcal{K}'$ contain more parameters, so that for some extra parameters in a set $\mathcal{K}''$, we have $\mathcal{K}' = \mathcal{K} \times \mathcal{K}''$. (Any parameter list $k' \in \mathcal{K}'$ has the form $k' = k \| k''$ where $k \in \mathcal{K}$, $k'' \in \mathcal{K}''$ and "$\|$" indicates concatenation.) In this type of extension of a parameter set, the injection can immerse (or even embed) a lower dimensional manifold into one of higher dimension.

Rather than extending it, an experimenter sometimes simplifies an experiment, which leads a theorist to simplify a parameter space, for instance by making some of the parameters on the list dependent on others. The simplest case is to fix the values of certain parameters of the list; then a simplified parameter set $\mathcal{K}'$ comes from deleting the parameters whose values are fixed, and the injection goes the other way $\mathcal{K}' \hookrightarrow \mathcal{K}$.

### A. Measure spaces

Experimental outcomes, whether viewed as discrete or on a continuum, are expressed in theory by points of some "outcome" set $\Omega$ equipped with a $\sigma$-algebra of measurable subsets. In the finite case of $n$ binary detectors, any point of $\Omega$ is a string of $n$ bits, the $j$-th bit being 1 if the $j$-detector fired and otherwise 0. In this case $\Omega$ consists of the $2^n$ possible strings of $n$ bits. By $\tilde{\Omega}$ we denote measurable subsets of $\Omega$, that is, the things that are assigned probabilities by a probability measure. Whenever $\Omega$ is a finite or countably infinite set, we can take $\tilde{\Omega}$ to be all the subsets of $\Omega$. To make our formulation more general we include uncountable outcome sets, such as the real line. When $\Omega$ is uncountable, then not all of its subsets can be measurable, so that the set $\tilde{\Omega}$ of measurable subsets of $\Omega$ is instead a $\sigma$-algebra [11].

An example of an outcome is a position as a number of meters; another example is momentum as a number of kg·m/s. We take each outcome—whether a binary number, an integer, a real number, or a list of real numbers—to be numerical; that is, we detach the outcome as a numerical entity from units, such as meters or seconds, required to interpret its significance. By this trick we avoid requiring distinct measure spaces for measurements of variables that are interpreted in distinct units.

A *probability measure* $\mu : \tilde{\Omega} \to [0, 1]$ is a positive measure of total measure 1; *i.e.*, it assigns to each $\omega \in \tilde{\Omega}$ a real number between 0 and 1, subject to axioms for unions and intersections [11]. Thus by $\mu$ we denote a probability *measure* while by $\mu(\omega)$ we denote a probability, meaning a number between 0 and 1, inclusive.

We denote the set of all probability measures on measurable sets $\tilde{\Omega}$ by $\text{PM}(\tilde{\Omega})$.

Describing some extensions of an experiment requires expanding the outcome set. For example, adding detectors corresponds to a new outcome set $\Omega' = \Omega \times \Omega''$, with a surjection (think of a projection) back to $\Omega$, $\xi : \Omega' \to \Omega$, such that if $\omega$ is measurable in $\Omega$, then $\xi^{-1}(\omega)$ is measurable in $\Omega'$; that is we have

$$(\forall \, \omega \in \tilde{\Omega}) \ \xi^{-1}(\omega) \in \tilde{\Omega}'. \tag{2.1}$$

The idea is that $\Omega$ lacks detail definable in $\Omega'$. Here is a simple example in which $\Omega$ models a single binary detector while $\Omega'$ models two detectors: let $\Omega = \{0, 1\}$, $\Omega' = \{00, 01, 10, 11\}$, and define $\xi$ by $\xi(ij) = i$ for $i, j \in \{0, 1\}$.

Thinking of $\Omega'$ as expressing an extension of an experiment by adding detectors, we expect a correspondence between a probability measure $\mu'$ on $\Omega'$ and $\mu$ on $\Omega$ under the surjection $\xi$:

$$(\forall \, \omega \in \tilde{\Omega}) \ \mu'(\xi^{-1}(\omega)) = \mu(\omega). \tag{2.2}$$

When this holds, we say the probability measure $\mu'$ *refines* the probability measure $\mu$ via the mapping $\xi$.

The set of probability measures $\text{PM}(\tilde{\Omega})$ can be equipped with a metric $D_{\tilde{\Omega}}$, which defines for any two measures $\mu, \lambda \in \text{PM}(\tilde{\Omega})$ a distance $D(\mu, \lambda)$. An option for a metric is the $L_1$ distance. For this, let $s$ denote a finite or countably infinite partition of the outcome set $\Omega$ into pairwise disjoint measurable subsets $\omega$. Let $S$ be the set of all such partitions. Then the $L_1$ distance between PMs $\mu$ and $\nu$ on $\tilde{\Omega}$ is defined by a supremum over partitions:

$$L_{1,\tilde{\Omega}}(\mu, \nu) = \frac{1}{2} \sup_{s \in S} \sum_{\omega \in s} |\mu(\omega) - \nu(\omega)|. \tag{2.3}$$

### B. Parametrized probability measures

Given a parameter space $\mathcal{K}$ and measurable sets $\tilde{\Omega}$, let a function that assigns to each $k \in \mathcal{K}$ a probability measure (PM) on $\tilde{\Omega}$ be called a *parametrized probability measure* (PPM) on the pair $(\mathcal{K}, \tilde{\Omega})$. Sometimes we look at a PPM as a a function $\mathcal{K} \to \text{PM}(\tilde{\Omega})$; alternatively we can see any PPM as a function $\mathcal{K} \times \tilde{\Omega} \to [0, 1]$ that assigns to $k$ and $\omega$ a probability. We often let $\mu^{(\cdot)}$ denote a PPM; then we denote the probability measure that it assigns to the parameter list $k$ by $\mu^{(k)}$, and we denote the probability that $\mu^{(k)}$ assigns to a measurable set $\omega$ by $\mu^{(k)}(\omega)$. In what follows it is necessary to distinguish among $\mu^{(\cdot)}$ as a PPM, $\mu^{(k)}$ as a PM, and $\mu^{(k)}(\omega)$ as a probability.

The mappings $\Xi : \mathcal{K} \hookrightarrow \mathcal{K}'$ and $\xi : \Omega' \to \Omega$ allow us to define the *envelopment* of one PPM by another. Suppose we have

$$(\forall\, k \in \mathcal{K})(\forall\, \omega \in \tilde{\Omega})\ \mu'^{(\Xi(k))}(\xi^{-1}(\omega)) = \mu^{(k)}(\omega). \tag{2.4}$$

In this case we say the PPM $\mu'^{(\cdot)}$ *envelops* the PPM $\mu^{(\cdot)}$ via $\Xi$ and $\xi$. Entailed in this envelopment is that $\mu'^{(\Xi(k))}$ refines $\mu^{(k)}$. Except when the injection $\Xi$ is surjective, there are elements of $\mathcal{K}' \setminus \Xi(\mathcal{K})$ that correspond to parameter values in $\mathcal{K}'$ that are unavailable in $\mathcal{K}$.

### C. Quantum models

The term *model* is employed in both physics and mathematics, but differently, and within physics two uses of *model* are worth distinguishing. One use in physics has to do with thinking about how to arrange the devices of an experiment: e.g., an experimenter may think of (model) the introduction of a polarization rotator as inserting a matrix (say as an element of the group SU(2)). In its other use in physics, which is of most concern here, "to model" is to use a PPM as a model of the relative frequencies of experimental outcomes in relation to experimentally controlled parameters. In contrast to both these uses, in mathematics, a model of a mathematical system of axioms consists of relations among entities defined in some other system of axioms.

PPMs serve as models in the sense of physics: they model relative frequencies of experimental outcomes in relation to experimentally controlled parameters. As we shall show shortly, given a probability measure, one can choose a density operation and a POVM that generate the probability measure (*via* Eq. (1.1)). Such a density operator and POVM constitute a model, in the sense of mathematics, of the given probability measure. Similarly dealing with probability measures, density operators, and the POVMs that are all functions of parameter lists, we arrive at *quantum models* as mathematical models of PPMs; we build up to their definition as follows:

1. By a POVM on measurable subsets $\tilde{\Omega}$ and a Hilbert space $\mathcal{H}$, we mean a function

    $M : \tilde{\Omega} \to \{\text{positive, bounded operators on } \mathcal{H}\}$, satisfying the condition: if $\{\omega_j\}$ is a finite or

countably infinite partition of the outcome set $\Omega$ into pairwise disjoint measurable subsets, then

$$\sum_j M(\omega_j) = \mathbf{1}_{\mathcal{H}}. \tag{2.5}$$

(See the discussion of "generalized observables" in [12].) We denote the set of all POVMs on $\tilde{\Omega}$ and $\mathcal{H}$ by POVM($\tilde{\Omega}, \mathcal{H}$).

2. By a density operator on $\mathcal{H}$ we mean a bounded, positive, trace-class operator having unit trace. Denote the set of all density operators on $\mathcal{H}$ by DensOps($\mathcal{H}$).

3. Suppose a PPM $\mu^{(\cdot)}$ on $(\mathcal{K}, \tilde{\Omega})$ is given. Adjusting our previous concept of a quantum model [4, 6], we now define a quantum model to be a mathematical model of a PPM, consisting of:

   (a) a separable Hilbert space $\mathcal{H}_\alpha$, along with

   (b) a *density-operator function* $\rho_\alpha^{(\cdot)} : \mathcal{K} \to \text{DensOps}(\mathcal{H}_\alpha)$, and

   (c) a *POVM-function* $M_\alpha^{(\cdot)} : \mathcal{K} \to \text{POVM}(\tilde{\Omega}, \mathcal{H})$,

   such that

$$(\forall k \in \mathcal{K})(\forall \omega \in \tilde{\Omega}) \ \text{Tr}[\rho_\alpha^{(k)} M_\alpha^{(k)}(\omega)] = \mu^{(k)}(\omega). \tag{2.6}$$

Given a PPM $\mu^{(\cdot)}$ on $(\mathcal{K}, \tilde{\Omega})$, we define the class of models that generate this $\mu^{(\cdot)}$ to be

$$\text{MODELS}(\mu^{(\cdot)}, \mathcal{K}, \tilde{\Omega}) = \{(\mathcal{H}, \rho^{(\cdot)}, M^{(\cdot)}) | (\forall \omega \in \tilde{\Omega}) \ \text{Tr}\left[\rho^{(\cdot)} M^{(\cdot)}(\omega)\right] = \mu^{(\cdot)}(\omega)\}, \tag{2.7}$$

with the understanding that for all $k \in \mathcal{K}$, $\rho^{(k)} \in \text{DensOps}(\mathcal{H})$ and $M^{(k)} \in \text{POVM}(\tilde{\Omega}, \mathcal{H})$. Note that the designation MODELS($\mu^{(\cdot)}, \mathcal{K}, \tilde{\Omega}$) specifies no particular Hilbert space, and this is important: distinct models having non-isomorphic Hilbert spaces can belong to the same class of models.

Usually, each parameter list $k$ is a concatenation of a sublist $k_{\text{prep}}$ for "state preparation" and a disjoint sublist $k_{\text{meas}}$ for "measurement": $k = k_{\text{prep}} \| k_{\text{meas}}$, in which case Eq. (2.6) specializes to

$$(\forall k \in \mathcal{K})(\forall \omega \in \tilde{\Omega}) \ \text{Tr}[\rho_\alpha^{(k_{\text{prep}})} M_\alpha^{(k_{\text{meas}})}(\omega)] = \mu^{(k)}(\omega). \tag{2.8}$$

## 3. MANY QUANTUM MODELS OF ANY PPM

When some PPM $\mu^{(\cdot)}$ on $(\mathcal{K}, \tilde{\Omega})$ has been abstracted from experimental data, one may want to choose a model to fit it. As proved in [6], this choice is highly non-unique. For instance, define the overlap of two density operators $\rho$ and $\rho'$ by

$$\text{Overlap}(\rho, \rho') \stackrel{\text{def}}{=} \text{Tr}[\rho^{1/2}\rho'^{1/2}]. \tag{3.1}$$

Then by Proposition 2 of [6], whenever MODELS$(\mu^{(\cdot)}, \mathcal{K}, \tilde{\Omega})$ contains a model $\alpha$ with the property that for some lists $k_1$ and $k_2$ the density operators $\rho_\alpha^{(k_1)}$ and $\rho_\alpha^{(k_2)}$ have a positive overlap, MODELS$(\mu^{(\cdot)}, \mathcal{K}, \tilde{\Omega})$ also contains a model $\beta$ for which the overlap of $\rho_\beta^{(k_1)}$ and $\rho_\beta^{(k_2)}$ is zero.

Another interesting question is this: does every PPM have a quantum model? At least when $\mathcal{K}$ is finite or countably infinite, the answer is "yes" as follows.

**Proposition**: Let $\mu^{(\cdot)}$ be any PPM on $(\mathcal{K}, \tilde{\Omega})$ with $\mathcal{K}$ finite or countably infinite; then MODELS$(\mu^{(\cdot)}, \mathcal{K}, \tilde{\Omega})$ is not empty. (3.2)

*Proof*: Without loss of generality, let $\mathcal{H}$ be a Hilbert space with an orthonormal basis $|j\rangle$, $j \in \mathcal{K}$. Let $\rho^{(k)} = |k\rangle\langle k|$ and define a POVM (independent of $k$) by

$$(\forall \omega \in \tilde{\Omega}) \ M(\omega) = \sum_{k' \in \mathcal{K}} \mu^{(k')}(\omega) |k'\rangle\langle k'|. \tag{3.3}$$

Then we have

$$(\forall k \in \mathcal{K})(\forall \omega \in \tilde{\Omega}) \ \text{Tr}[\rho^{(k)} M(\omega)] = \mu^{(k)}(\omega). \quad \square \tag{3.4}$$

Suppose now that $\mathcal{K} \subset \mathcal{K}_{\text{prep}} \times \mathcal{K}_{\text{meas}}$, so that each list of parameters splits into two sublists: $k = k_{\text{prep}} \| k_{\text{meas}}$. Does every PPM on such a $\mathcal{K}$ have a quantum model that respects this split, in the sense that $k_{\text{prep}} \in \mathcal{K}_{\text{prep}}$ specifies a density operator while $k_{\text{meas}} \in \mathcal{K}_{\text{meas}}$ specifies a POVM?

**Proposition**: Let $\mu^{(\cdot)}$ be any PPM on $(\mathcal{K}, \tilde{\Omega})$ with $\mathcal{K} = \mathcal{K}_{\text{prep}} \times \mathcal{K}_{\text{meas}}$; if $\mathcal{K}_{\text{prep}}$ is finite or countably infinite, MODELS$(\mu^{(\cdot)}, \mathcal{K}, \tilde{\Omega})$ contains a model that respects the splitting of $\mathcal{K}$; that is, there exists a Hilbert space $\mathcal{H}$, a density-operator function
$$\rho^{(\cdot)}: \mathcal{K}_{\text{prep}} \to \text{DensOps}(\mathcal{H}), \text{ and a POVM-function } M^{(\cdot)}: \mathcal{K}_{\text{meas}} \to \text{POVM}(\tilde{\Omega}, \mathcal{H}) \quad (3.5)$$
satisfying
$$(\forall\, k \in \mathcal{K})(\forall\, \omega \in \tilde{\Omega})\ \ \text{Tr}[\rho^{(k_{\text{prep}})} M^{(k_{\text{meas}})}(\omega)] = \mu^{(k)}(\omega).$$

*Proof*: Let $\mathcal{H}$ be a Hilbert space with an orthonormal basis $|j\rangle$, $j \in \mathcal{K}_{\text{prep}}$. For any $k_{\text{prep}} \in \mathcal{K}_{\text{prep}}$, let $\rho^{(k_{\text{prep}})} = |k_{\text{prep}}\rangle\langle k_{\text{prep}}|$ and define POVM-function on $\mathcal{K}_{\text{meas}}$ by

$$(\forall\, \omega \in \tilde{\Omega})\ \ M^{(k_{\text{meas}})}(\omega) = \sum_{k_{\text{prep}}} \mu^{(k_{\text{prep}} \| k_{\text{meas}})}(\omega)\, |k_{\text{prep}}\rangle\langle k_{\text{prep}}|. \quad (3.6)$$

(Because of the sum, $M^{(k_{\text{meas}})}$ is independent of $k_{\text{prep}}$.) Then we have

$$(\forall\, k \in \mathcal{K})(\forall\, \omega \in \tilde{\Omega})\ \ \text{Tr}[\rho^{(k_{\text{prep}})} M^{(k_{\text{meas}})}(\omega)] = \mu^{(k)}(\omega). \quad \square \quad (3.7)$$

**Remark**: If one adds the requirement that the detection operators are tensor products, as in [5], then no such proposition holds.

**Remark**: Belonging to the same class MODELS$(\mu^{(\cdot)}, \mathcal{K}, \tilde{\Omega})$ is a much looser relation among models than is unitary equivalence. For two models $\alpha$ and $\beta$ of a given class, the Hilbert spaces $\mathcal{H}_\alpha$ and $\mathcal{H}_\beta$ can differ in dimension, and the inner products of pure states of model $\alpha$ need not match those of model $\beta$ [6]. In addition, one encounters models $\alpha$ and $\beta$ of the same class with unequal von Neumann entropies: for some $k \in \mathcal{K}$, one has

$$S(\rho_\alpha^{(k)}) \neq S(\rho_\beta^{(k)}), \quad (3.8)$$

where $S$ denotes the von Neumann entropy of a density operator

$$S(\rho) \stackrel{\text{def}}{=} -\text{Tr}[\rho \log_2(\rho)], \quad (3.9)$$

with the understanding that $S(\rho)$ can be infinite. In particular, there are cases of two models in the same class with $S(\rho_\alpha^{(k)})$ infinite while $S(\rho_\beta^{(k)})$ is finite.

**Remark**: In choosing a model within a given class, the fit to probabilities offers no guidance, for all models in the class exhibit the same probability measures over $\mathcal{K}$.

**Remark**: Rather than choosing a density-operator function and a POVM-function to fit a given PPM, often a theorist goes the other way by thinking first in terms of quantum states (for instance, wave functions) and operators. These can be put in the form of a pair $(\rho^{(\cdot)}, M^{(\cdot)})$—a density-operator function and POVM-function—which generate a PPM via Eq. (2.8). Indeed, this is the path of prediction. The theorist can invoke concepts of particles, leading to quantum states for electrons, photons, *etc*. When a PPM calculated from particle-associated states matches a subsequently performed experiment, the PPM is confirmed as a prediction. But what does this tell us about the particles and their associated quantum states? One confirms that the particles lead to a PPM that fits the experiment, but as proved in [6], there are lots of other configurations of particles and their quantum states that serve as alternative models of the same PPM. See Ref. [13] for an example of the merit of attending to alternative hypotheses in evaluating claims for the security of quantum key distribution.

### A. Application of non-uniqueness of models to the Holevo bound

Invoked in quantum communication [14] is a model $\alpha$ of a quantum channel, in which Alice with probability $p_\alpha^{(i)}$ prepares a state $\rho_\alpha^{(i)}$ transmitted to a receiver Bob described by a fixed POVM $M_\alpha$, as illustrated in Fig. 1. Bob's outcome set $\Omega_\alpha$ is finite or countably infinite and consists of disjoint measurable events $j = 1, 2, \ldots$. Model $\alpha$ generates the conditional probability of Bob's outcome $j$ given that Alice prepares $\rho_\alpha^{(i)}$ as

$$\mu_\alpha^{(i)}(j) = \text{Tr}[\rho_\alpha^{(i)} M_\alpha(j)]. \tag{3.10}$$

We denote the mutual information between $A$ (for Alice) and $B$ (for Bob) ascribed by model $\alpha$ by $I_\alpha(A, B)$. The definition of this mutual information makes use of Shannon entropy, defined for any discrete probability measure $\mu$ (alternately written $\mu(\cdot)$) by

$$H(\mu(\cdot)) \stackrel{\text{def}}{=} -\sum_j \mu(j) \log_2[\mu(j)]. \tag{3.11}$$

Note that if $\mu^{(i)}$ is a probability measure, so is $\sum_i p^{(i)} \mu^{(i)}$. From the definition of the mutual information [15] we have

$$\begin{aligned} I_\alpha(A, B) &= H\left(\sum_i p_\alpha^{(i)} \mu_\alpha^{(i)}(\cdot)\right) - \sum_i p_\alpha^{(i)} H\left(\mu_\alpha^{(i)}(\cdot)\right) \\ &= H\left(\sum_i p_\alpha^{(i)} \text{Tr}[\rho_\alpha^{(i)} M_\alpha(\cdot)]\right) - \sum_i p_\alpha^{(i)} H\left(\text{Tr}[\rho_\alpha^{(i)} M_\alpha(\cdot)]\right). \end{aligned} \tag{3.12}$$

For mutual information derived from density operators and POVMs, Holevo [16] showed an upper bound:

$$I(A, B) \le \chi(A, B) \stackrel{\text{def}}{=} S\left(\sum_i p^{(i)} \rho^{(i)}\right) - \sum_i p^{(i)} S\left(\rho^{(i)}\right), \tag{3.13}$$

where $S$ is the von Neumann entropy defined in Eq. (3.9). This bound is useful in many cases, but when frequency is accounted for [17], the von Neumann entropy can be arbitrarily large, indeed infinite, in which case the Holevo bound tells us nothing. Notice, however, that in Eq. (3.12) for mutual information, the density operators enter only as factors in products with the POVM detection operators. For this reason, for any models $\alpha$ and $\beta$, we have

$$\mu_\alpha^{(\cdot)} = \mu_\beta^{(\cdot)} \Rightarrow I_\alpha(A, B) = I_\beta(A, B). \tag{3.14}$$

From this follows the tighter bound on mutual information:

$$I_\alpha(A, B) \le \min_{\beta \in [\alpha]} \chi_\beta(A, B), \tag{3.15}$$

where we write $[\alpha]$ as shorthand for $\text{MODELS}(\mu_\alpha^{(\cdot)}, \mathcal{K}, \tilde{\Omega})$.

### B. Example from quantum cryptography of an enveloping PPM

The envelopment of one PPM by another has application to showing a likely vulnerability to a widely discussed design for quantum key distribution (QKD) [18]. In its simplest form, QKD calls for Alice to transmit key bits along with some extra bits to Bob. Each bit is carried by a signal modeled as a quantum state, with the promise that if Alice and Bob compare notes on the extra bits (over a public channel), any significant eavesdropping must induce errors that Alice and Bob's comparison will reveal to them. The promise is of security against *undetected* eavesdropping.

For quantum states as light states, BB84 [19] posits a sequence of trials, at each of which Alice prepares one of four possible single-photon light states, chosen at random. In the original design, this is expressed by a model $\alpha$ with a Hilbert space $\mathcal{H}_\alpha$ taken as a two-dimensional real vector space; each bit is expressed by a density operator $\rho_\alpha^{(\theta)} = |\theta\rangle\langle\theta|$, where, for orthonormal basis states $|x\rangle$ and $|y\rangle$,

$$|\theta\rangle = \cos\theta |x\rangle + \sin\theta |y\rangle; \tag{3.16}$$

here the four allowed values of $\theta$ are given by

$$\theta \in \mathcal{K}_{\alpha,\text{prep}} = \{0, \pi/4, \pi/2, 3\pi/4\}. \tag{3.17}$$

For present purposes we consider only vulnerabilities in the face of an intercept-resend attack [18], in which an eavesdropper Eve, interposed between Alice and Bob, measures the state that Alice prepares and then sends to Bob whatever she chooses. For pointing to a vulnerability present in some implementations of BB84, it suffices to assume that (1) the transmission is lossless; (2) Eve's measurements are by a pair of orthogonally oriented detectors of perfect efficiency and zero dark count; and (3) Eve can freely choose the orientation of her detectors, so that one detector responds optimally to light states polarized at any angle $\theta'$ of Eve's choice, while the other detector responds optimally to the light state polarized at $\theta \pm \pi/2$.

Under these assumptions, Eve's outcome set $\Omega_\alpha = \{0, 1\}$, with 1 for the event of detection for polarization angle $\theta'$ and 0 for the event of detection at a polarization angle perpendicular to $\theta'$. Eve's measurement is expressed by the projection-valued POVM-function defined on $\theta' \in \mathcal{K}_{\alpha,\text{meas}} = S^1$, where $S^1$ denotes the unit circle $[0 \leq \theta' < 2\pi)$:

$$\begin{aligned} M_\alpha^{(\theta')}(1) &= |\theta'\rangle\langle\theta'|, \\ M_\alpha^{(\theta')}(0) &= 1 - |\theta'\rangle\langle\theta'|. \end{aligned} \tag{3.18}$$

This POVM-function combines with the density-operator function to generate a PPM defined by

$$\mu_\alpha^{(\theta,\theta')}(1) = \text{Tr}[\rho^{(\theta)} M^{(\theta')}] = |\langle\theta|\theta'\rangle|^2 = \cos 2(\theta - \theta'),$$
$$\mu_\alpha^{(\theta,\theta')}(0) = 1 - \mu_\alpha^{(\theta,\theta')}(1) = 1 - |\langle\theta|\theta'\rangle|^2 = \sin 2(\theta - \theta'). \quad (3.19)$$

The claim of BB84 for security against undetected eavesdropping depends on positive inner products such as $\langle\theta = 0|\theta = \pi/4\rangle = 2^{-1/2}$. Given these inner products as they enter the PPM of Eq. (3.19), there appears no way for an eavesdropper to detect with certainty what state Alice transmits. (For the secret of how Bob manages to extract a good key, see Ref. [19].)

Now comes the *implementation* of BB84 with lasers and optical fibers and so on. In this, one encounters challenges unexpressed by model $\alpha$, and meeting these challenges demands models that envelop model $\alpha$. Here is a case in point. BB84 calls for pulse-to-pulse changes in polarization, and these are technically difficult to arrange. To evade this difficulty, one implementation employs four separate lasers, each transmitting one of the four needed polarizations. The complication is that this implementation demands precisely matched laser frequencies, an issue invisible to model $\alpha$, which leaves the frequency spectrum of the light pulses unexpressed. To think about light frequency, we envelop $\mu_\alpha^{(\cdot)}$ by a finer-grained PPM $\mu_\beta^{(\cdot)}$ produced by a model $\beta$ that explicitly expresses frequency spectra. In model $\beta$ (which invokes a Hilbert space $\mathcal{H}_\beta$ of infinite dimension) a single-photon state linearly polarized at an angle $\theta$ is expressed [17] by

$$|\theta, f\rangle = \int d\omega f(\omega) a_\theta^\dagger(\omega)|0\rangle, \quad (3.20)$$

where $a_\theta^\dagger(\omega)$ is a creation operator for light at the angular frequency $\omega$, polarized at angle $\theta$ [17]). (Apologies for using the variable $\omega$ here for angular frequency of a light spectrum; we depend on context to distinguish this use from our main use of $\omega$ as a measurable subset of an outcome set.) Correspondingly, we have $\rho_\beta^{(\theta,f)} = |\theta, f\rangle\langle\theta, f|$. The outcome set is the same: $\Omega_\beta = \{0, 1\}$; however, the space of preparation parameters is a lot bigger:

$$\mathcal{K}_{\beta,\text{prep}} = S^1 \times \{f | f \in L^2(R), \int_I d\omega |f(\omega)|^2 = 1\}, \quad (3.21)$$

where by $L^2(R)$ we mean square-integrable functions on the real line. Suppose we limit attention to a class of functions on $R$ for which we have a complete set of orthonormal functions $f_j$. We then have a POVM-function defined on $(\theta', j)$, $j \in N^+ = \{1, 2, \ldots\}$ by

$$M_\beta^{(\theta',j)}(1) = |\theta', f_j\rangle\langle\theta', f_j|,$$
$$M_\beta^{(\theta',j)}(0) = |\theta'_\perp, f_j\rangle\langle\theta'_\perp, f_j|, \quad (3.22)$$

where $\theta_\perp$ is an angle perpendicular to $\theta$ in the plane of linear polarization. This gets us to a measurement-parameter space

$$\mathcal{K}_{\beta,\text{meas}} = S^1 \times N^+. \quad (3.23)$$

The POVM-function defined in Eq. (3.22) combines with the density-operator function to generate a PPM on $\mathcal{K}_{\beta,\text{prep}} \times \mathcal{K}_{\beta,\text{meas}}$, defined by

$$\mu_\beta^{(\theta,f,\theta',j)}(1) = \text{Tr}[\rho^{(\theta,f)}M^{(\theta',j)}] = |\langle\theta, f|\theta', f_j\rangle|^2$$
$$= \cos^2(\theta - \theta')\left|\int_I d\omega f^*(\omega)f_j(\omega)\right|^2,$$
$$\mu_\beta^{(\theta,f,\theta',j)}(0) = |\langle\theta, f|\theta'_\perp, f_j\rangle|^2$$
$$= \sin^2(\theta - \theta')\left|\int_I d\omega f^*(\omega)f_j(\omega)\right|^2. \quad (3.24)$$

Now $\mu_\beta^{(\cdot)}$ envelops $\mu_\alpha^{(\cdot)}$ via the identification of the measure spaces and, for example, the injection $\Xi(\theta) = (\theta, f_1)$. But $\mathcal{K}_\beta$ contains lots of elements outside the image of this injection. For example, if Alice's lasers of 1.5 nm nominal wavelength and pulses of 1 ns duration are imperfectly matched in frequency by one part in $10^{-5}$, then the inner products that in model $\alpha$ appear to be $2^{-1/2}$ can all become essentially zero, as expressed in model $\beta$ by virtue of the integrals in Eq. (3.24). Under this circumstance any expectation of security against an undetected intercept-resend eavesdropping attack is misplaced. (For a different way in which the envelopment of one PPM by another can challenge claims of cryptographic security, see Ref. [20].)

### 4. PARAMETER SPACES AND PPM'S ASSOCIATED WITH ENTANGLED STATES

Ten years after Born discussed them [21], interest in non-factorizable quantum states jumped when Einstein, Podolsky, and Rosen (EPR) [22] examined the theory of their measurement.

Decades later interest in such states, now called *entangled*, jumped again when Bell displayed certain correlations visible in what in our words is a PPM associated with an entangled state [23]—correlations that violate inequalities satisfied by local hidden-variable theories. Later still Gisin showed that any entangled pure state can be associated with a PPM that violates generalized Bell inequalities [7, 24].

From a few well-known examples in which quantum states strongly violate Bell inequalities, we extract parameter spaces and PPMs to see patterns in places where we had never before thought to look. While a wider variety of patterns invite future attention, the patterns displayed here are primarily topological. Of interest are the topological features of the parameter spaces and also features of what we call a *level set* of the PPM over a parameter space $\mathcal{K}$, meaning a subset $[k] \subset \mathcal{K}$ over which the probability measure assigned by the PPM takes a particular value; more formally, any PPM $\mu(\cdot)$ over $\mathcal{K}$ partitions $\mathcal{K}$ into level sets $[k]$ by

$$[k] \stackrel{\text{def}}{=} \{k' \in \mathcal{K} | \mu^{(k')} = \mu^{(k)}\}. \tag{4.1}$$

### A. Formulation

We limit our attention to cases of a pure state $|\psi\rangle$, corresponding to density operator $\rho = |\psi\rangle\langle\psi|$, so that for any POVM $M$ on the Hilbert space $\mathcal{H}$ in which $|\psi\rangle$ resides, we have

$$\text{Tr}[\rho M(\omega)] = \langle\psi|M(\omega)|\psi\rangle. \tag{4.2}$$

Given any tensor-product factorization of the Hilbert space $\mathcal{H}$ into

$$\mathcal{H} = \mathcal{H}_A \otimes \mathcal{H}_B, \tag{4.3}$$

one conventionally calls a pure state $|\psi\rangle$ *un*entangled (with respect to this factorization) if

$$(\exists\, |\psi_A\rangle \in \mathcal{H}_A,\ |\psi_B\rangle \in \mathcal{H}_B)\ |\psi\rangle = |\psi_A\rangle \otimes |\psi_B\rangle); \tag{4.4}$$

otherwise it is said to be *entangled*.

## B. EPR

Reporting on the theory of measuring what we now call an entangled state, EPR speak of two particles, 1 and 2, that interact for a limited time and then separate, after which they are expressed by a wave function of a sort that is now called *entangled*. EPR discuss measurements made in two parts, one part at one location for particle 1, the other part at a location widely separated from the first for particle 2. Each local measurement involved a choice of parameter, corresponding in the language here of a choice of $k \in \mathcal{K}_{\text{meas}}$. For their example of an entangled state, EPR showed that the variance of a probability measure for outcome events at one location varied with the choice of measurement made at the other location.

Following the EPR notation, we set aside for the moment our use of the letters $A$ and $B$. Note also that instead of POVMs, which came later, EPR use hermitian operators to express measurements, so that rather than dealing with probability measures they deal with only expectation values and variances. They write $A$ for the momentum operator for a particle 1, $B$ for the position operator of particle 1, $P$ for the momentum operator for particle 2, and $Q$ for the position operator of particle 2. The issue is the circumstances under which the variance in the probability measure for outcomes is narrow or wide. They show that the choice of whether $A$ or $B$ is measured at 1 changes which of $P$ or $Q$ yields a narrow variance at 2.

Rephrasing this in the language of this chapter, rather than limiting attention to expectation values, variances, and possibly higher moments associated with hermitian operators for "quantities," we note that to each such operator, via the spectral decomposition [25], one arrives at a POVM. In this way the EPR notion of a "quantity that has with certainty the value" $x$ translates to a probability measure having a sharp peak that concentrates its mass at or near the value $x$.

From our viewpoint, EPR deal with discrete parameters $(k_1, k_2)$ with $k_1 \in \{A, B\} \cong \{0, 1\}$ and $k_2 \in \{P, Q\} \cong \{0, 1\}$. Taking the outcome sets to be $\Omega_1 = \Omega_2 = R$ (real numbers), regardless of whether the outcome pertains to meters or to momentum, their state and measurement operators generate a PPM over the four combinations $(k_1, k_2)$ as shown in Fig. 2. (Note that EPR use $\delta$- functions, leading to relative probabilities; in the rest of our work, we smear any $\delta$-functions to get plain probabilities.)

### C. Generalized Bell inequalities and their violation

As with EPR, Bell's and many later studies of entangled states [24] also invoke measurement operators, mostly projections, which we view as a specialization of POVMs. Viewed this way, these studies show features of PPMs derived from separating a measurement into separate measurements in two locations A and B. The separation is expressed in part by:

1. a parameter space $\mathcal{K}_{\text{meas}} \subset \mathcal{K}_A \times \mathcal{K}_B$;

2. an outcome space $\Omega = \Omega_A \times \Omega_B$, each cartesian-product factor having its own measurable subsets, $\widetilde{\Omega}_A$ and $\widetilde{\Omega}_B$, respectively.

This partitioning of parameter lists and of measurable sets implies a PPM of the form

$$\mu^{(k_A,k_B)}(\omega_A, \omega_B).$$

In addition, the PPM $\mu^{(\cdot,\cdot)}$ inherits restrictions that follow from the form of any of the models involving entangled states that generate it. In any such model the separation into locations A and B finds additional expression as

1. a Hilbert space $\mathcal{H} = \mathcal{H}_A \otimes \mathcal{H}_B$;

2. a POVM-function with POVMs $M^{(k_A,k_B)} = M_A^{(k_A)} \otimes M_B^{(k_B)}$, where $M_A^{(k_A)} \in \text{POVM}(\widetilde{\Omega}_A, \mathcal{H}_A)$ and $M_B^{(k_B)} \in (\widetilde{\Omega}_B, \mathcal{H}_B)$.

This factorization imposes on the PPM the form

$$\mu^{(k_A,k_B)}(\omega_A, \omega_B) = \langle \psi | M_A^{(k_A)}(\omega_A) \otimes M_B^{(k_B)}(\omega_B) | \psi \rangle. \tag{4.5}$$

(Because we focus on a single state, we hold $k_{\text{prep}}$ fixed and omit writing it.)

For any entangled state there exists a set of four unentangled POVMs for which the corresponding four probability measures violate a generalized Bell inequality [7]. Combined with the proof that the PPM that displays violations can be generated by disparate quantum models, this makes the PPM $\mu^{(\cdot,\cdot)}$ itself an interesting object to study.

### D. No-signaling condition

The assumption of factorization of POVMs has an implication for the probability measures involving them; namely, the well known "no-signaling condition" is satisfied by the marginal probabilities:

$$(\forall\, \omega_A \in \widetilde{\Omega}_A)\ \mu^{(k_A,\cancel{k_B},\ldots)}(\omega_A) \stackrel{\text{def}}{=} \mu^{(k_A,\cancel{k_B},\ldots)}(\omega_A, \Omega_B) \text{ is independent of } k_B,$$

$$(\forall\, \omega_B \in \widetilde{\Omega}_B)\ \mu^{(\cancel{k_A},k_B,\ldots)}(\omega_B) \stackrel{\text{def}}{=} \mu^{(\cancel{k_A},k_B,\ldots)}(\Omega_A, \omega_B) \text{ is independent of } k_A, \quad (4.6)$$

where we put a slash through a parameter sublist to which the indicated parametrized probabilities are insensitive.

### E. Case study of a toroidal parameter space

To within questions of detector inefficiencies and how to model them, a series of experiments demonstrates clear violations of Bell inequalities. We focus on one discussed in [26] and depicted in Fig. 3. A light source radiates entangled single-photon states in two oppositely directed beams, $A$ and $B$. Each beam impinges on a measuring assembly consisting of a polarizing beam splitter followed by two sensitive light detectors, which we call 1 and 2. For purposes of appreciating the theory of entangled states, we pretend that the detectors have no dark counts and perfect efficiency and that the transmission involves no losses. Then, following [27], we arrive at a model for which the outcome set $\Omega_A$ consists of just the two points $\{1, 2\}$ that correspond to one and only one detector firing, and the same for $\Omega_B$.

Suppose for a moment that each assembly can be rotated to any angle $\theta/2$ around the beam axis; this would correspond to detector 1 responding to light linearly polarized at angle $\theta/2$ and detector 2 responding to light lineally polarized at angle $(\theta + \pi)/2$ leading to a family of POVMs $M_A^{(\theta_A)}$ on the outcome set $\Omega_A = \{1, 2\}$, and a similar family for $B$.

The most convenient expression of this can be found in [27], where one deals with a model light state

$$|\psi\rangle = \frac{1}{\sqrt{2}} (|x_A\rangle|x_B\rangle + |y_A\rangle|y_B\rangle). \tag{4.7}$$

For this state one considers (projection-valued) POVMs, with the POVM $M^{(k_A)}$ (where $k_A = \theta_A$) defined by

$$M_A^{(\theta_A)}(1) = \left(\cos\frac{\theta_A}{2}|x\rangle + \sin\frac{\theta_A}{2}|y\rangle\right)\left(\cos\frac{\theta_A}{2}\langle x| + \sin\frac{\theta_A}{2}\langle y|\right),$$

$$M_A^{(\theta_A)}(2) = 1 - M_A^{(\theta_A)}(1) = \left(\sin\frac{\theta_A}{2}|x\rangle - \cos\frac{\theta_A}{2}|y\rangle\right)\left(\sin\frac{\theta_A}{2}\langle x| - \cos\frac{\theta_A}{2}\langle y|\right). \tag{4.8}$$

Replacing $A$ by $B$ throughout yields $M_B^{(\theta_B)}$. Via Eq. (1.1), this defines a PPM on a torus $\mathcal{K} = S^1 \times S^1$, with coordinates $(\theta_A, \theta_B)$, corresponding to detectors set to respond to linearly polarized light with a polarization angle $\theta_A/2$ at $A$ and $\theta_B/2$ at $B$. Thus,

$$\mu^{(\theta_A,\theta_B)}(1_A, 1_B) = \mu^{(\theta_A,\theta_B)}(2_A, 2_B) = \frac{1}{4}[1 + \cos(\theta_A - \theta_B)],$$

$$\mu^{(\theta_A,\theta_B)}(1_A, 2_B) = \mu^{(\theta_A,\theta_B)}(2_A, 1_B) = \frac{1}{4}[1 - \cos(\theta_A - \theta_B)]. \tag{4.9}$$

A kind of "correlation" is defined on $\mathcal{K}$ in terms of this PPM by

$$E(\theta_A, \theta_B) = \mu^{(\theta_A,\theta_B)}(1_A, 1_B) + \mu^{(\theta_A,\theta_B)}(2_A, 2_B) - \mu^{(\theta_A,\theta_B)}(1_A, 2_B) - \mu^{(\theta_A,\theta_B)}(2_A, 1_B)$$

$$= \cos(\theta_A - \theta_B). \tag{4.10}$$

The generalized Bell inequality at issue involves the correlation $E$ evaluated at four pairs of values of the angles:

$$-2 \leq S_{\text{Bell}} \leq 2, \tag{4.11}$$

for

$$S_{\text{Bell}} \stackrel{\text{def}}{=} E(\theta_A, \theta_B) - E(\theta_A, \theta_B') + E(\theta_A', \theta_B) + E(\theta_A', \theta_B'). \tag{4.12}$$

As shown at the top in Fig. 4, the parameter space with points $(\theta_A, \theta_B)$ is a torus. Skinning the surface of the donut and stretching it out, we get the square area in Fig. 4, which shows the contours of constant $\mu$, namely the lines $\theta_A - \theta_B = $ const. The level sets of $\mu^{(\cdot)}$ as a function on $\mathcal{K}$ are thus closed loops that wind once around the torus. The figure also shows values of $\theta_A$ and $\theta_B$ for which $S_{\text{Bell}}$ maximally violates the inequality.

### F. Allowing for elliptical polarization: SO(3)

Linear polarization is needlessly restrictive. To display the drama ensuing from general elliptical polarization, we need a different light state. By inserting a suitable element in the path toward detector $B$, one can change the light modeled by the $|\psi\rangle$ defined in Eq. (4.7) into something modeled by a singlet state

$$|\psi'\rangle = \frac{1}{\sqrt{2}} (|x_A\rangle|y_B\rangle - |y_A\rangle|x_B\rangle), \qquad (4.13)$$

where now we view $|x\rangle$ and $|y\rangle$ as orthonormal vectors in the complex vector space $\mathbb{C}^2$. The advantage of the state $|\psi\rangle$ is its invariance under the same SU(2) transformation applied to both the $A$- and $B$-factor states. We now expand the parameter space $\mathcal{K}_{\text{meas}}$ to allow for detections not just of linearly polarized light but light of arbitrary elliptical polarization. With the same assumptions as above about perfectly efficient detectors and no dark counts, one has a detector assembly modeled by a POVM with an additional parameter $\phi$:

$$\begin{aligned}
M_A^{(\theta_A,\phi_A)}(1) &= \left(\cos\frac{\theta_A}{2}|x\rangle + e^{i\phi_A}\sin\frac{\theta_A}{2}|y\rangle\right)\left(\cos\frac{\theta_A}{2}\langle x| + e^{-i\phi_A}\sin\frac{\theta_A}{2}\langle y|\right), \\
M_A^{(\theta_A,\phi_A)}(2) &= 1 - M_A^{(\theta_A,\phi_A)}(1) \\
&= \left(\sin\frac{\theta_A}{2}|x\rangle - e^{i\phi_A}\cos\frac{\theta_A}{2}|y\rangle\right)\left(\sin\frac{\theta_A}{2}\langle x| - e^{-i\phi_A}\cos\frac{\theta_A}{2}\langle y|\right), \quad (4.14)
\end{aligned}$$

along with the same expression in which $A$ is replaced by $B$ throughout. Here the ranges of the parameters are $0 \leq \theta_{A,B} \leq \pi$ and $0 \leq \phi < 2\pi$.

The general unit ray for an elliptically polarized state, represented by $\cos(\theta_A/2)|x\rangle + e^{i\phi_A}\sin(\theta_A/2)|y\rangle$, corresponds to a point on the Poincaré sphere: $\phi$ corresponds to longitude while $\theta$

corresponds to latitude measured down from the North pole. By this mapping, the orthogonal state goes to the polar opposite point on the sphere. The selection of a POVM thus corresponds to the selection of two points, with coordinates $(\theta_A, \phi_A)$ on a sphere for $A$, and $(\theta_B, \phi_B)$ on a sphere for $B$. Instead of the torus $S^1 \times S^1$ as the parameter space when polarization is constrained to be linear, for elliptical polarization we find a parameter space that is a product of two spheres:

$$\mathcal{K}_{\text{meas}} = S^2 \times S^2. \tag{4.15}$$

It is straightforward to calculate the PPM on $S^2 \times S^2$ for this case:

$$\begin{aligned}
\mu^{(\theta_A,\phi_A,\theta_B,\phi_B)}(1_A, 1_B) &= \mu^{(\theta_A,\phi_A,\theta_B,\phi_B)}(2_A, 2_B) \\
&= \frac{1}{4}[1 - \cos\theta_A \cos\theta_B - \sin\theta_A \sin\theta_B \cos(\phi_A - \phi_B)], \\
\mu^{(\theta_A,\phi_A,\theta_B,\phi_B)}(1_A, 2_B) &= \mu^{(\theta_A,\phi_A,\theta_B,\phi_B)}(2_A, 1_B) \\
&= \frac{1}{4}[1 + \cos\theta_A \cos\theta_B + \sin\theta_A \sin\theta_B \cos(\phi_A - \phi_B)].
\end{aligned} \tag{4.16}$$

On recognizing a little spherical trigonometry, one notices the following about the level sets of probability measures of the PPM defined by Eq. (4.16) on $\mathcal{K}_{\text{meas}} = S^2 \times S^2$. We can view $S^2 \times S^2$ as a space in which each element is a pair of points, $(\theta_A, \phi_A)$ and $(\theta_B, \phi_B)$ on one unit sphere. Let $\zeta$ be the angle between the unit vectors to these points; $0 \leq \zeta \leq \pi$. Then we have

$$\cos\zeta = \cos\theta_A \cos\theta_B + \sin\theta_A \sin\theta_B \cos(\phi_A - \phi_B), \tag{4.17}$$

so that Eq. (4.16) simplifies to

$$\begin{aligned}
\mu^{(\theta_A,\phi_A,\theta_B,\phi_B)}(1_A, 1_B) &= \mu^{(\theta_A,\phi_A,\theta_B,\phi_B)}(2_A, 2_B) = \frac{1}{4}[1 - \cos\zeta], \\
\mu^{(\theta_A,\phi_A,\theta_B,\phi_B)}(1_A, 2_B) &= \mu^{(\theta_A,\phi_A,\theta_B,\phi_B)}(2_A, 1_B) = \frac{1}{4}[1 + \cos\zeta].
\end{aligned} \tag{4.18}$$

This PPM maximally violates a Bell inequality.

In the previous case, limited to linear polarization, all the level sets of probability measures on the torus $S^1 \times S^1$ were, apart from location on the torus, the same one-dimensional loop; but here with the parameter space $S^2 \times S^2$ we have some variety. Any pair of points defines an angle $\zeta$ between the rays from the origin of the sphere to the points, and any level set for probability measures over $\mathcal{K} = S^2 \times S^2$ is the locus of all pairs having any fixed angle $\zeta$ between them, $0 \le \zeta \le 2\pi$. For the open interval $0 < \zeta < \pi$, this level set is a manifold of dimension 3 obtained as follows. Starting from any pair of points $(\theta_A, \phi_A)$ and $(\theta_B, \phi_B)$ on the unit sphere separated by angle $\zeta$ in the open interval, every other pair separated by $\zeta$ can be reached by one and only one rotation acting on the starting pair, so the level set is the orbit under an effective action of the rotation group SO(3). The manifold of all these pairs—the level set—is isomorphic to SO(3) as a Lie group, which, like the circle, is connected but not simply connected. In contrast for $\zeta = 0$ (where $(\theta_A, \phi_A) = (\theta_B, \phi_B)$) or for $\zeta = \pi$ (where they are polar images of one another), SO(3) has an isotropy group of SO(2) and instead of the three-dimensional manifold of SO(3), the level set is just the simply connected two-dimensional manifold, the sphere $S^2$.

### G. Property of local reach

The PPMs of the preceding two subsections dealing with violations of Bell inequalities have a special property (not shared by some other PPMs derived from entangled states): By holding $k_A$ fixed at any one value and varying $k_B$ (or *vice versa*) one covers the whole space of measures. For example, the probability measures involved in the simplest model above of polarization-entangled light are functions of $(k_A - k_B)$, where $k_A \equiv \theta_A$ and $k_B \equiv \theta_B$ are single numbers that express the angles to which polarizing filters are set [3]. More generally we will say that a PPM $\mu^{(\cdot,\cdot)}$ has the property of "local reach" if:

$$(\forall\, k_A, k_B, k'_B)(\exists\, k'_A)\ \mu^{(k_A, k_B)} = \mu^{(k'_A, k'_B)}, \tag{4.19}$$

$$(\forall\, k_A, k_B, k'_A)(\exists\, k'_B)\ \mu^{(k_A, k_B)} = \mu^{(k'_A, k'_B)}. \tag{4.20}$$

**Proposition**: Consider any PPM $\mu^{(\cdot)}$ on some $\mathcal{K} = \mathcal{K}_{\text{prep}} \times \mathcal{K}_{\text{meas}}$, with $\mathcal{K}_{\text{meas}} = \mathcal{K}_A \times \mathcal{K}_B$, and suppose that this PPM satisfies the no-signaling condition and has the property of local reach; then the marginal probabilities for all its probability measures are invariant over $\mathcal{K}_{\text{meas}}$. (4.21)

*Proof*: Evaluate Eq. (4.20) for the probability of $(\omega_A, \Omega_B)$ to get

$$(\forall k_A, k_B, k'_A)(\exists k'_B)(\forall \omega_A \in \widetilde{\Omega}_A) \ \mu^{(k_A, k_B)}(\omega_A, \Omega_B) = \mu^{(k'_A, k'_B)}(\omega_A, \Omega_B). \quad (4.22)$$

By the assumed no-signaling equation, the marginal probability is insensitive to $k_B$, whence we have

$$(\forall k_A, k'_A)(\forall \omega_A \in \widetilde{\Omega}_A) \ \mu^{(k_A, \cancel{k_B})}(\omega_A, \Omega_B) = \mu^{(k'_A, \cancel{k_B})}(\omega_A, \Omega_B), \quad (4.23)$$

where the slash through $k_B$ indicates that its value makes no difference. Hence we have that $\mu^{(k_A, \cancel{k_B})}(\omega_A, \Omega_B)$ depends neither on $k_A$ nor on $k_B$; thus we can write $\mu^{(\cancel{k_A}, \cancel{k_B})}(\omega_A, \Omega_B)$. The same argument holds for $\mu^{(\cancel{k_A}, \cancel{k_B})}(\Omega_A, \omega_B)$. □

### H. Allowing for light frequency, etc.

As discussed in the example of quantum cryptography in Sec. 3 B, light involves more degrees of freedom than just polarization. Not only do light pulses come with frequency spectra, implying an infinite-dimensional Hilbert space, but models can posit states of more than one photon [17]. As we saw in Sec. 3 B, under various special assumptions, such as frequency-independent detectors, models invoking the infinite dimensional spaces provide PPMs that envelop those discussed in the preceding two subsections. These finer-grained models provide a much bigger class of PPMs, the characterization of which awaits future exploration.

### I. Metrical properties

Bell inequalities express metrical properties of $\mu^{(\cdot)}$. For instance, as it works in the example of a toroidal parameter space, Eq. (4.11) expresses an $L_1$-distance between $\mu^{(\theta_A, \theta_B)}$ and $\mu^{(\theta_A, \theta'_B)}$ along with more complicated properties that quantify how much and in what ways a PPM varies both with $k_{\text{meas}}$ and with $\omega$. It is worth noting that the hidden-variable model discussed in [27] has the same toroidal parameter space as discussed in Sec. 4 E, but with contours such that no violation of the Bell inequality (4.11) is possible.

As is well known in quantum decision theory [14], a metric on PM($\tilde{\Omega}$) also provides a basis for deciding between one value of a parameter and another value. For example, consider the situation a little more general than that sketched in Fig. 1. This situation, which arises not only when Alice and Bob are communicating parties, but also when they are opponents, is described by a PPM $\mu^{(\cdot)}$ over some $(\mathcal{K}, \widetilde{\Omega}_B)$ with $\mathcal{K} = \mathcal{K}_A \times \mathcal{K}_B$. Alice chooses $k_A \in \mathcal{K}_A$, Bob chooses $k_B \in \mathcal{K}_B$. Then Bob's probability of outcome $\omega_B \in \widetilde{\Omega}_B$ is $\mu^{(k_A, k_B)}(\omega_B)$. How well can Bob distinguish Alice's parameter values $k_A$ and $k'_A$? We say that these values are $\epsilon$-separable for Bob's choice of measurement $k_B$ when

$$\epsilon \leq D_{\tilde{\Omega}}\left(\mu^{(k_A, k_B)}, \mu^{(k'_A, k_B)}\right). \tag{4.24}$$

As an avenue for future work, we are interested in the case in which the preparation parameters to be decided include spacetime coordinates.

Also for future exploration is the issue of PPMs that are "close together." To quantify the difference between two such PPMs over a given parameter set and given measurable sets, one defines a metric on them. A plausible metric is based on the value of $k$ at which the two probability measures most differ. That is, given a distance $D_{\tilde{\Omega}}$ on probability measures, we get a distance between two PPMs. For $\mu^{(\cdot)}$ and $\mu'^{(\cdot)}$ over $\mathcal{K} \times \tilde{\Omega}$ a distance between them can be defined by:

$$\mathcal{D}\left(\mu^{(\cdot)}, \mu'^{(\cdot)}\right) \stackrel{\text{def}}{=} \sup_k D_{\tilde{\Omega}}\left(\mu^{(k)}, \mu'^{(k)}\right). \tag{4.25}$$

Such a distance is useful in comparing a PPM from a model against a second PPM extracted from experimental relative frequencies.

## Acknowledgments

We thank Tai Tsun Wu, Howard Brandt, and Samuel Lomonaco for vigorous discussions of the linking of equations to experiments. JM thanks Horace Yuen for the chance to talk about an earlier version of this paper during a visit to Northwestern University. He also thanks Prem Kumar for helpful discussions, David Mumford for an illuminating discussion of $S^3 \times S^3$, and Martin Jaspan for the invitation to participate in experiments in quantum key distribution.

# Figure Captions

Figure 1.  Quantum communication channel (binary case).

Figure 2.  EPR parametrized probabilities.

Figure 3.  Experiment to detect polarization-entangled photons; PBS = polarizing beam splitter.

Figure 4.  Parameter space $S^1 \times S^1$ and contours of constant probability measure $E$ on torus with marked points $(\theta_A, \theta_B) = \{(-3\pi/8, -\pi/8), (\pi/8, -\pi/8), (\pi/8, 3\pi/8), (-3\pi/8, 3\pi/8)\}$ for a maximal violation of the Bell inequality.

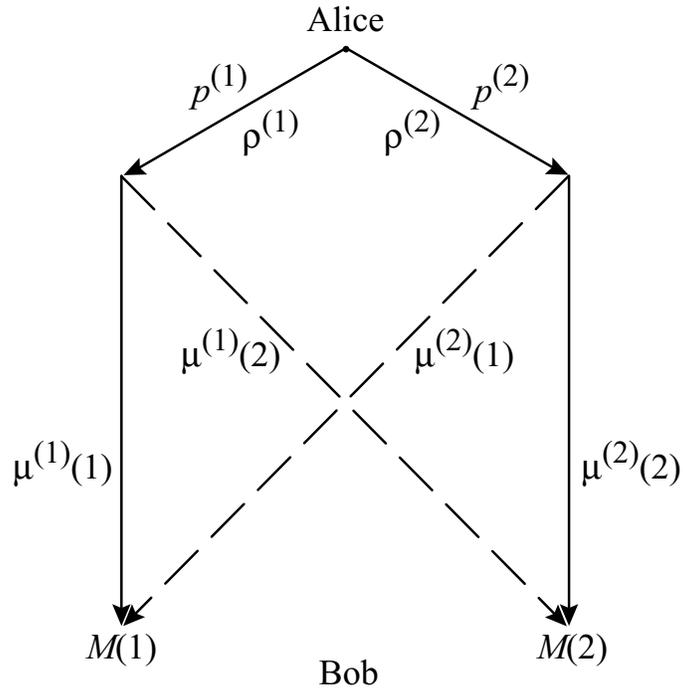

FIG. 1: Quantum communication channel (binary case).

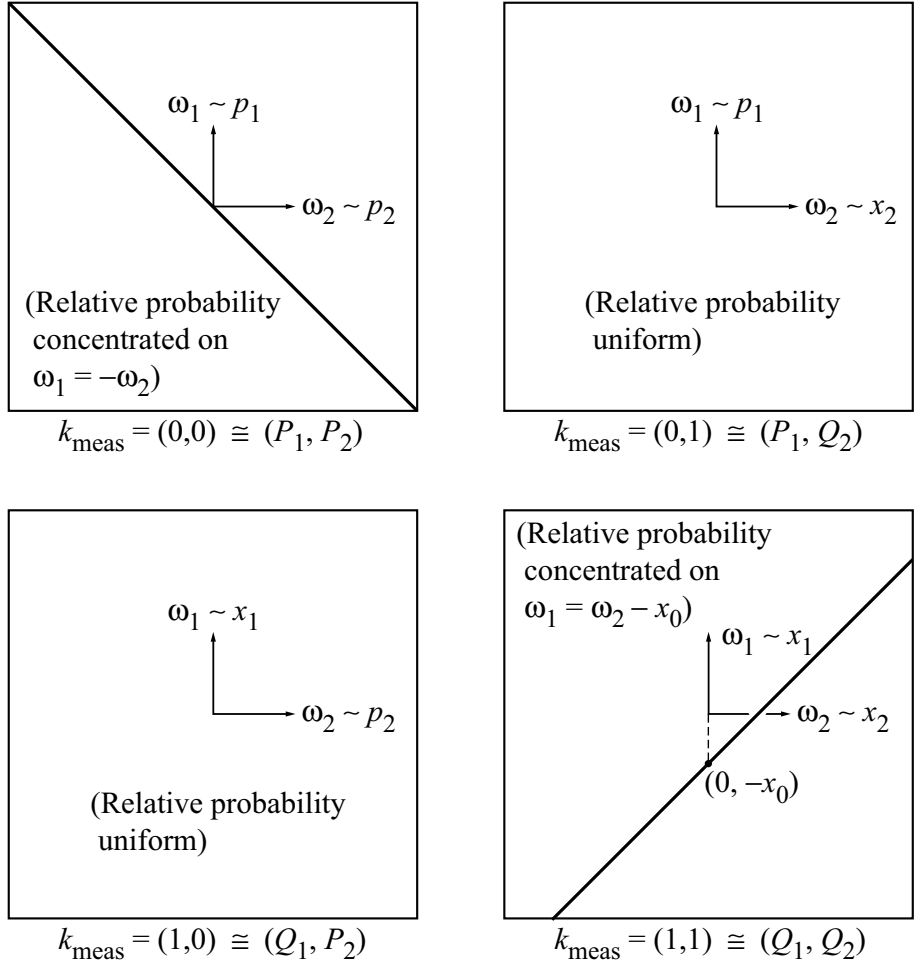

FIG. 2: EPR parametrized probabilities.

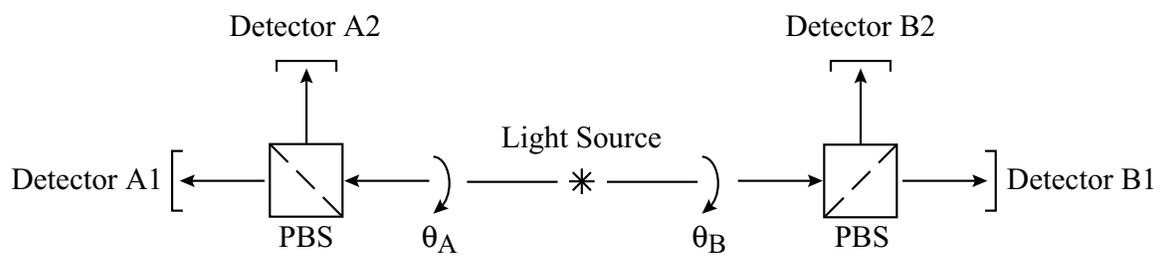

FIG. 3: Experiment to detect polarization-entangled photons. (PBS = polarizing beam splitter.)

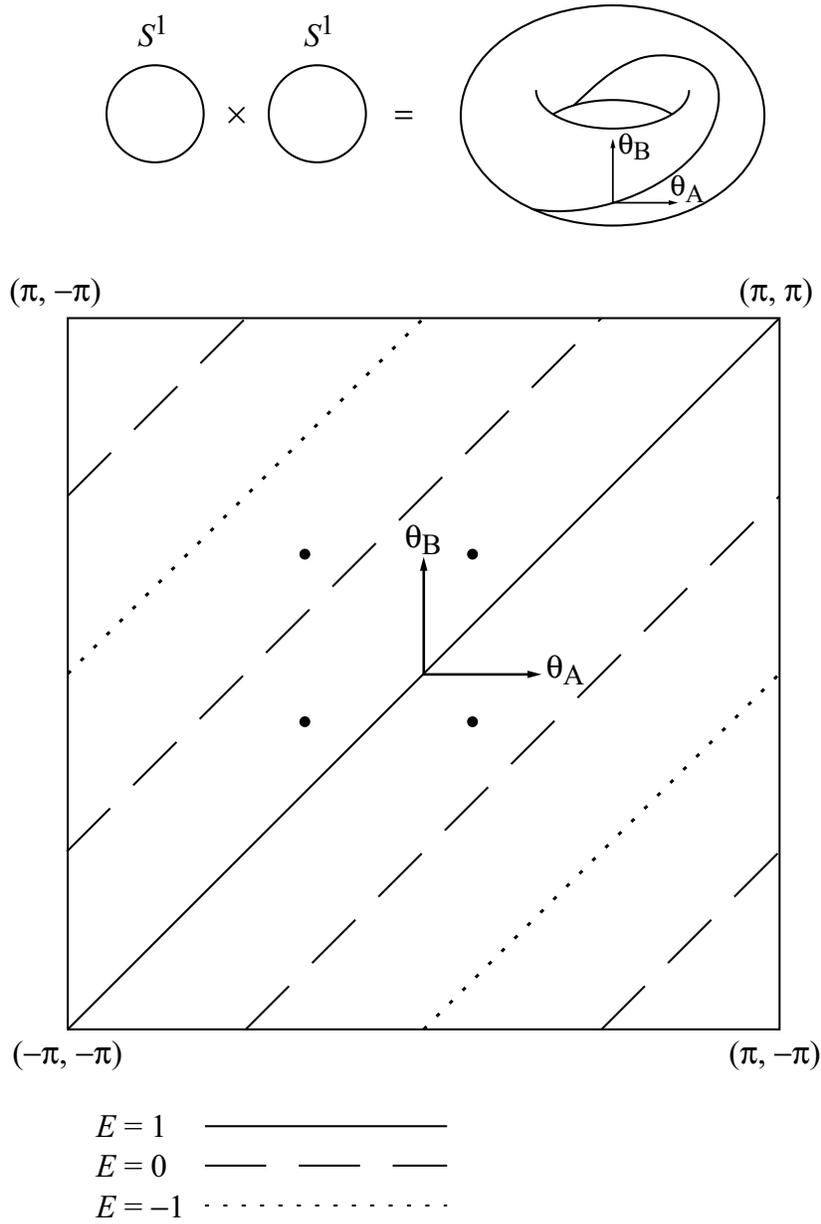

FIG. 4: Parameter space $S^1 \times S^1$ and contours of constant probability measure $E$ on torus with marked points $(\theta_A, \theta_B) = \{(-3\pi/8, -\pi/8), (\pi/8, -\pi/8), (\pi/8, 3\pi/8), (-3\pi/8, 3\pi/8)\}$ for a maximal violation of the Bell inequality.